\documentclass[sigconf,screen]{acmart}
\AtBeginDocument{%
  }

 \usepackage{booktabs}
\usepackage{hyperref}
\usepackage{makecell}

\usepackage{pifont}

\usepackage{xcolor} % 必须有

\newcommand{\qj}[1]{\textcolor{black}{#1}}

\newcommand{\final}[1]{\textcolor{black}{#1}}

  \usepackage{changepage}

\copyrightyear{2026}
\acmYear{2026}
\setcopyright{cc}
\setcctype{by}
\acmConference[CHI '26]{Proceedings of the 2026 CHI Conference on Human Factors in Computing Systems}{April 13--17, 2026}{Barcelona, Spain}
\acmBooktitle{Proceedings of the 2026 CHI Conference on Human Factors in Computing Systems (CHI '26), April 13--17, 2026, Barcelona, Spain}
\acmPrice{}
\acmDOI{10.1145/3772318.3790299}
\acmISBN{979-8-4007-2278-3/2026/04}

\begin{document}

\title[Presence as Predictors of Basic Psychological Need Satisfaction]{Social, Spatial, and Self-Presence as Predictors of Basic Psychological Need Satisfaction in Social Virtual Reality}

\author{Qijia Chen}
\orcid{0000-0003-1038-5461}
\affiliation{%
\department{Department of Computer Science}
  \institution{University of Helsinki}
  \city{Helsinki}
   \country{Finland}
}
\email{qijia.chen@helsinki.fi}

\author{Andrea Bellucci}
\orcid{0000-0003-4035-5271}
\affiliation{%
\department{Department of Computer Science and Engineering}
  \institution{Universidad Carlos III de Madrid}
  \city{Leganés}
  \country{Spain}
  }
\email{abellucc@inf.uc3m.es}

\author{Giulio Jacucci}
\orcid{0000-0002-9185-7928}
\affiliation{
\department{Department of Computer Science}
  \institution{University of Helsinki}
  \city{Helsinki}
  \country{Finland}
  }
\email{giulio.jacucci@helsinki.fi }

\renewcommand{\shortauthors}{Qijia et al.}

\begin{abstract}
\qj{
Extensive research has examined presence and basic psychological needs \final{(drawing on Self-Determination Theory)} in digital media. While prior work offers hints of potential connections, we lack a systematic account of whether and how distinct presence dimensions map onto the basic needs \final{of autonomy, competence, and relatedness}.
We surveyed 301 social VR users and analyzed using Structural Equation Modeling. Results show that social presence predicts all three needs, while self-presence predicts competence and relatedness, and spatial presence shows no direct or moderating effects.} Gender and age moderated these relationships: women benefited more from social presence for autonomy and relatedness, men from self- and spatial presence for competence and autonomy, and younger users showed stronger associations between social presence and relatedness, and between self-presence and autonomy. These findings position presence as a motivational mechanism shaped by demographic factors. The results offer theoretical insights and practical implications for designing inclusive, need-supportive multiuser VR environments.

%Extensive research has examined presence and basic psychological needs in digital media, yet we still lack a systematic account of how distinct presence dimensions map onto these needs in social VR. Survey data from 301 social VR users analyzed with structural equation modeling show that social presence predicts all three needs, self-presence selectively predicts competence and relatedness, and spatial presence shows no direct or moderating effects. Gender and age further shape these patterns—women benefit more from social presence for autonomy and relatedness, men from self- and spatial presence for competence and autonomy, and younger users show stronger links between social presence and relatedness and between self-presence and autonomy—positioning presence as a motivational mechanism shaped by demographic factors and informing the design of inclusive, need-supportive multiuser VR environments. 

\end{abstract}



\maketitle

\section{Introduction}
Social Virtual Reality (VR) has rapidly emerged as an important medium for digital communication and shared experience~\cite{mcveigh2018s}. Unlike earlier platforms that relied on text or video, social VR foregrounds embodied interaction through avatars, spatialized audio, and shared 3D environments. People meet, play, collaborate, and perform together in these spaces~\cite{Hugging5, 9417636}. \qj{They often report} a heightened sense of being with others and being within a virtual world~\citep{Sykownik2023VR, chen2024d}. This makes social VR a timely setting to examine not just how people perceive immersive technologies, but how these environments shape psychological experiences such as motivation, well-being, and connection.

A defining feature of social VR is presence, the subjective sense of “being there” in a mediated environment~\cite{steuer1992defining}. Decades of VR research shown that presence is multidimensional~\cite{ lee2004presence, riva2014interacting}. It can be understood as spatial presence (feeling located in the virtual environment), social presence (feeling with and attended to by others), and self-presence (identifying with one’s avatar or virtual body) \citep{lee2004presence}. %Crucially, meta-analytic and survey evidence demonstrates that higher display fidelity (e.g., wider fields of view, lower latency, greater resolution) enhances sensory realism and stabilizes sensorimotor contingencies, while embodiment features such as visuomotor congruence, precise hand/eye tracking, and a stable first-person perspective reinforce body ownership and agency~\cite{slater2016enhancing, kilteni2012sense, Cummings2016ImmersionPresence}. 
Crucially, \qj{meta-analytic evidence} shows that higher display fidelity (e.g., wider fields of view, lower latency) enhances sensory realism and stabilizes sensorimotor contingencies~\cite{Cummings2016ImmersionPresence}. \qj{In parallel, studies demonstrate that} factors like visuomotor congruence and precise hand/eye tracking reinforce body ownership and agency~\cite{slater2016enhancing, kilteni2012sense}.
Together, these advances strengthen presence and its subdimensions. These perceptual changes, in turn, carry cognitive and behavioral outcomes, a stronger presence is associated with improved task performance and motor efficiency, more fluid social coordination and rapport~\cite{slater2016enhancing, skarbez2017survey}. In short, when fidelity is high and sensorimotor contingencies are consistent, users tend to act more freely, perform more effectively, and feel more connected. \qj{These changes} are precisely the pathways through which presence shapes downstream outcomes.
%---precisely the pathways through which presence can shape downstream outcomes. %Presence thus structures how users interpret and act in immersive systems, and carries downstream effects for performance, collaboration, and social bonding~\cite{barreda2022psychological}. 

Thus, presence, beyond the perceptual illusion of ``being there,'' may also function as a motivational mechanism. Self-Determination Theory (SDT) \cite{ryan2017self} posits that the quality of motivation and well-being depends on satisfying three basic needs: autonomy (volition and choice), competence (effectiveness and mastery), and relatedness (connection to others). %When these needs are fulfilled, individuals display greater intrinsic motivation, engagement, and well-being; when they are thwarted, motivation and psychological functioning suffer~\cite{deci2000and, ryan2017self}.  %These basic needs are robust predictors of intrinsic motivation and well-being across contexts such as games, education, and social 
If presence shapes how freely users can act (autonomy), how capable they feel (competence), and how connected they become (relatedness), then different presence dimensions should map systematically onto these needs. 
However, despite extensive work on SDT in digital environments \cite{deci2000and, ryan2017self} \qj{and in HCI in general \citep{tyack2024self,ballou2022self}}, we still lack systematic exploration on how social, spatial, and self-presence are associated with psychological need satisfaction in social VR and VR more broadly. 
Clarifying these links matters because it positions presence (with its dimensions) not just as perceptual fidelity, but as a motivational pathway that may \qj{support or undermine} engagement, sustained use, and well-being. 
%\cite{ballou2022self,tyack2024self}

Moreover, presence effects may not be universal. Prior work suggests that individual factors such as gender and age may shape how immersive experiences are processed and their associated outcomes ~\citep{felnhofer2012virtual, melo2018presence,dilanchian2021pilot}. Similarly, spatial presence may amplify the effects of social or self-presence by making interactions feel more embodied and actionable~\cite{slater1997framework}. Understanding these contingencies is important for HCI, since it allows us to explain for whom presence supports motivation, and how VR environments can be designed to accommodate diverse user needs.
Guided by the perspectives, we ask:

\begin{quote}
\textbf{RQ1.} How do social, spatial, and self-presence predict users’ perceived autonomy, competence, and relatedness in social VR? 
\end{quote}

\begin{quote}
\textbf{RQ2.} \qj{\textbf{(a)} Does spatial presence moderate the associations between social and self-presence and psychological need 
satisfaction? \textbf{(b)} Do user characteristics, specifically gender and age, moderate the associations between presence dimensions and psychological need satisfaction?}
\end{quote}

To address these questions, we conducted a survey study of 301 social VR users. Using Structural Equation Modeling (SEM) and multi-group analyses, we systematically examined both the direct and moderating effects of presence on SDT outcomes. This study makes three key contributions to the literature:

\begin{itemize}
    \item \qj{The study links social, self-, and spatial presence to the needs for autonomy, competence, and relatedness. This framing positions} presence as a motivational, not just a perceptual, construct. It advances a multidimensional understanding of presence. %within the framework of Self-Determination Theory.

     \item The study demonstrates that gender and age significantly moderate how presence dimensions translate into need satisfaction. It reveals user-specific patterns in how immersive experiences are interpreted and internalized. This contributes to presence theory by demonstrating that presence's motivational impact is not universal, but shaped by user characteristics.

      \item We offer a theory-informed basis for designing social VR environments that accommodate diverse user needs. It also identifies directions for future research on intersectionality and mediators. %\ab{We highlight the importance of strengthening the presence dimensions that most directly support motivation, and of tailoring social VR features so that different user groups can translate presence into autonomy, competence, and relatedness.}

\end{itemize}

\section{Background}
\qj{This section presents relevant concepts that underpin our study. We first introduce social VR and its immersive affordances, followed by an exploration of the concept of presence and its dimensions. We then present the core psychological needs of Self-Determination Theory.}

%Taken together, these characteristics position social VR as a particularly promising context for studying how immersive, embodied social interaction relates to users’ underlying psychological experiences. Rather than tightly scripted laboratory exposures, social VR platforms provide a naturalistic setting in which people voluntarily engage, return, and build relationships over time. This makes them well suited for examining how core motivational constructs—such as basic psychological needs and sense of presence—are supported or thwarted in everyday, user-driven virtual encounters, which we elaborate in the following subsections.

\subsection{Social Virtual Reality}

Social VR is an emerging form of digital environment where users meet and interact in immersive, three-dimensional spaces \cite{Sykownik2023VR, maloney2021social}.
Social VR leverages real-time 3D avatars, spatial audio, and interactive virtual spaces to cultivate a heightened sense of presence among participants \cite{bailenson2018experience, tanenbaum2020make}. The adoption of advanced VR hardware (including motion tracking and eye-tracking) \qj{further enhances users’ sense of embodiment and “being there” with others in a virtual space \cite{piitulainen2022vibing}.}

\qj{Social VR platforms, including VRChat, Meta Horizon Worlds, and Rec Room, support a wide range of everyday and playful activities \citep{9417636,chen2025mirror}. Broader reviews highlight that social VR applications span co-watching, collaboration, live events, and other shared experiences \citep{SocialJie, li2023socialvr}. Users can hang out in virtual lounges, attend concerts and live events, play games, or engage in more mundane shared practices such as dancing \citep{piitulainen2022vibing}, drinking \citep{chen2024d}, or even sleeping together \cite{yin2023drifting}. Prior research shows that these environments can afford rich social experiences, from casual encounters to close-knit communities \citep{maloney2020falling}, and can serve as spaces for emotional support and social experimentation \cite{Hugging5, 9417636}.} \qj{Complementing these accounts of everyday practices, prior work has examined concrete co-present activities in social VR. For example, researchers have explored the role of proxemics in shaping group dynamics \citep{williamson2021proxemics} and the use of photo-sharing to foster social bonding \citep{li2019measuring}.}  

\qj{The immersive and embodied qualities of social VR also appear to foster deeper forms of social engagement than many non-immersive media. Studies suggest that social VR can facilitate personal disclosure and intimacy beyond what is typically reported in video chat or voice-only communication \cite{Maloney2020}. Such interactions potentially strengthen interpersonal trust and perceived closeness \cite{Hugging5}.} %Users often describe interactions in social VR as more “present” and emotionally resonant than comparable interactions on flat-screen platforms.
\qj{Together, social VR platforms are settings in which users’ everyday social experiences, emotions, and identities are negotiated through immersive, embodied interaction.}

\subsection{Sense of Presence}

Presence is commonly defined as the perceptual illusion of “being there” in a mediated environment \citep{steuer1992defining}.
It is a foundational construct in VR~\cite{steuer1992defining}, digital games~\cite{park2009understanding}, and computer-mediated environments~\cite{riva2014interacting}. 
Although conceptualizations vary, scholars generally agree that presence is a sensory-driven, automatically generated feeling of immersion \cite{lee2004presence,wirth2007process}. \qj{It captures the extent to which users experience a mediated environment as the place where they are currently located and acting.
Presence has been conceptualized as a multidimensional phenomenon~\cite{riva2014interacting, lee2004presence}. 
Following Lee~\cite{lee2004presence}, we distinguish between spatial presence, social presence, and self-presence. 
We adopt the distinction because it is widely used in VR studies and aligns with how presence is typically operationalized in recent social VR research \citep{barreda2022psychological,van2023feelings}.} %and it 

Spatial presence refers to the sensation of being physically “inside” a virtual or mediated environment rather than merely observing it on a screen~\cite{wirth2007process}. \qj{Social presence captures the extent to which individuals perceive others as co-located, attentive, and available for interaction in a mediated space~\cite{biocca2003toward,lowenthal2010social}. It reflects the felt quality of interpersonal encounters and social connection~\cite{biocca2003toward,skarbez2017survey}. 
Self-presence denotes the degree to which users experience their virtual representation as an extension of their own body and identity~\cite{ratan2013self,kilteni2012sense}. It indicates how strongly they identify with and embody their digital self.}

\qj{Technological advances have made it easier to elicit these forms of presence in VR. On the display side, higher resolution, wider fields of view, and low-latency head tracking increase sensory realism and make visual and vestibular cues more consistent \citep{Cummings2016ImmersionPresence}. On the interaction side, accurate hand, body tracking, spatialised audio, and haptic feedback stabilize sensorimotor contingencies~\cite{Cummings2016ImmersionPresence,slater2016enhancing}. Avatar embodiment features such as full-body animation, and a stable first-person viewpoint further strengthen feelings of body ownership and self-location in the virtual body~\cite{kilteni2012sense,Gonzalez}. \qj{Research on responsive, human-like virtual agent motion likewise supports these processes by enhancing spatial plausibility and sensorimotor coherence \cite{allbeck2007controlling}}.
These developments have been linked to higher ratings of spatial presence, co-presence with other users, and identification with avatars across games, training, and telecommunication contexts~\cite{Cummings2016ImmersionPresence,skarbez2017survey,riva2014interacting}.} 

\qj{\final{Higher levels of spatial, social, and self presence, in turn, have measurable consequences for how people think and act.} Experiments show that \final{higher spatial presence} is associated with more efficient navigation and motor ability in training, better memory for spatial layouts, and more successful transfer of learned skills to the physical world~\cite{slater2016enhancing,wei2025towards}. In multi-user scenarios, \final{higher social presence} predicts increased conversational involvement, smoother turn-taking, greater trust, and stronger feelings of rapport~\cite{biocca2003toward,skarbez2017survey}. Avatar embodiment has also been tied to changes in self-perception and behaviour, such as increased self-efficacy or prosocial intentions when embodying empowered avatars~\cite{ratan2013self}. When users feel more ``there'' they are more likely to treat virtual situations as consequential, act more freely, and experience interactions as more meaningful.}

\subsection{Self-Determination Theory and the Basic Psychological Needs}

Self-Determination Theory (SDT) is a well-established psychological framework that explores human motivation, well-being, and behavior through the lens of innate psychological needs \cite{deci2012self, deci2000and, deci2013intrinsic}. \qj{At its core, Basic Psychological Needs Theory posits that three fundamental needs (i.e., autonomy, competence, and relatedness) must be satisfied for people to experience high-quality motivation and psychological growth \cite{ryan2017self}.}

Autonomy refers to the need to experience volition and self-endorsed behavior, where individuals feel in control of their actions and choices. When autonomy is satisfied, individuals perceive their behaviors as aligned with their values and interests. %, leading to greater motivation and engagement.
Competence is the need to feel effective and capable in one's interactions with the environment. It involves mastering tasks, overcoming challenges, and developing skills contributing to a sense of achievement and self-efficacy.
Relatedness represents the need to feel connected to others, experience a sense of belonging, and maintain meaningful relationships \cite{ryan2017self, vansteenkiste2020basic}. %It is satisfied when individuals experience reciprocal care and support from social groups \cite{ryan2017self, vansteenkiste2020basic}.
These needs are universal, applying across cultures, contexts, and life stages. Their satisfaction predicts higher-quality, more persistent motivation. Their frustration undermines intrinsic motivation, leading to disengagement, stress, and potentially psychopathology \cite{ryan2010self, vansteenkiste2013psychological}.

\qj{Research on basic psychological needs has extended to various domains, including education, work, health, sports, and digital environments. In digital environments, need satisfaction is closely linked to sustained engagement, intrinsic motivation, and positive user experiences, while need frustration can lead to withdrawal or negative affect~\cite{formosa2022need,masur2014interplay}. Online platforms heavily rely on ongoing voluntary use, understanding how their affordances support or hinder autonomy, competence, and relatedness has become increasingly important~\cite{zhang2015understanding,tyack2024self}.}

%\ab{A growing body of work applies SDT to VR and metaverse experiences, showing its value for understanding motivation in immersive media \citep{Tsai2025ImmersiveLearningSDT, Green2024SDTStoryVR}. However, most of this research focuses on specific, purpose-built applications and uses need satisfaction to predict outcomes such as enjoyment \citep{Tsai2025ImmersiveLearningSDT} or learning \citep{Gim2023MetaverseSDT}. They do not examine how core experiential constructs of VR (e.g., presence) relate to basic psychological needs.} 

\qj{Recent work has also begun to apply SDT explicitly to VR and metaverse experiences \citep{Tsai2025ImmersiveLearningSDT, Green2024SDTStoryVR}. For example, \citet{Gim2023MetaverseSDT} examined how autonomy, competence, and relatedness in VR-based metaverse courses contribute to learner satisfaction, \citet{Green2024SDTStoryVR} analysed how IDEA-themed\footnote{\qj{IDEA stands for inclusive, diverse, equitable, and accessible.}} VR storytelling supports or diminishes basic psychological needs. These studies demonstrate that SDT is a useful framework for understanding motivation in immersive media. However, they focus on specific, purpose-built applications and typically model need satisfaction as a predictor of enjoyment, learning, or evaluation outcomes; or use VR broadly as a context. They do not examine how core experiential constructs of VR (e.g., presence) predict basic psychological needs.}

\section{\qj{Rationale for Research Questions and Hypotheses}}
\begin{figure}[h]
    \centering
    \includegraphics[width=.93\linewidth]{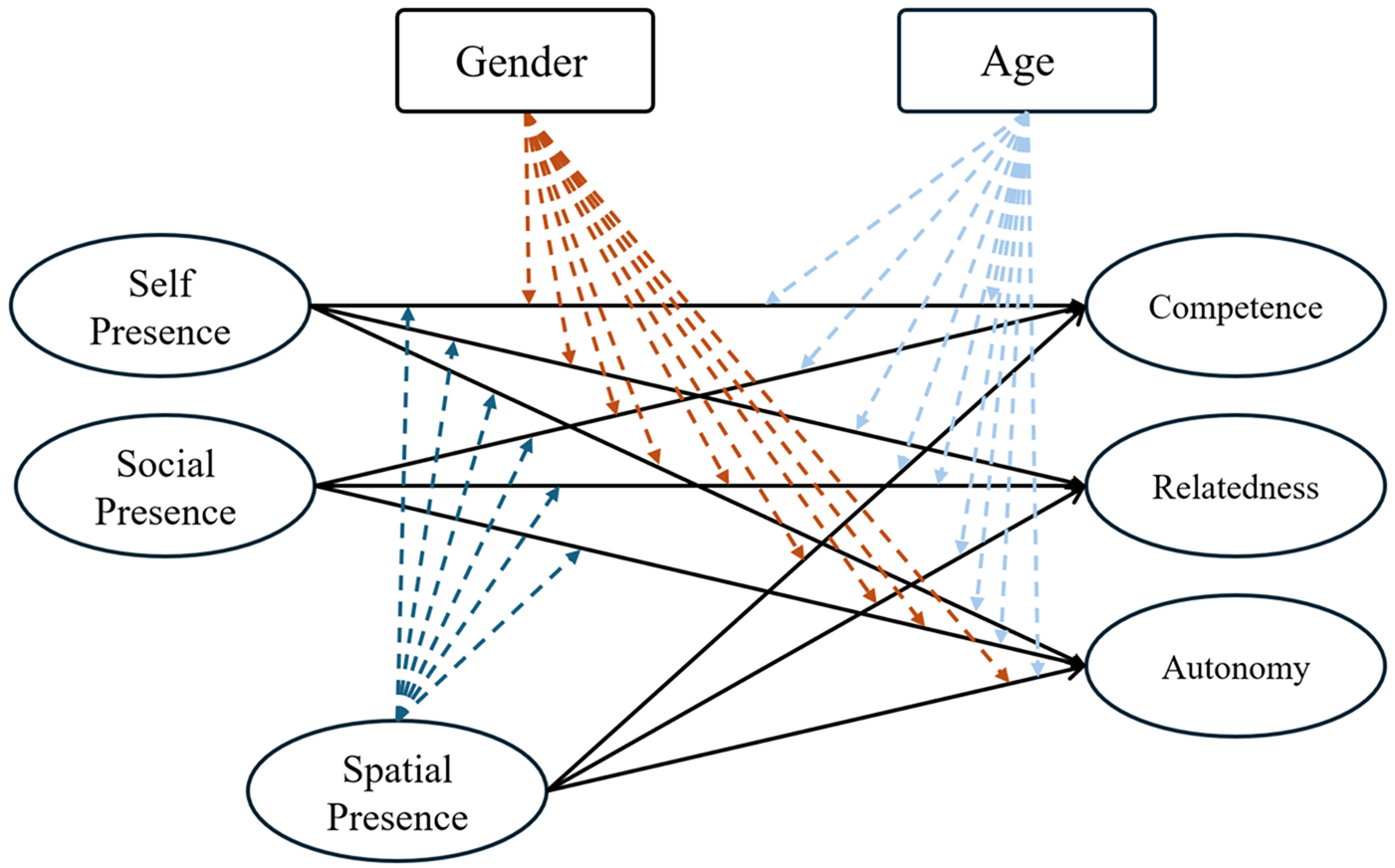}
\caption{\qj{\textbf{Hypothesised structural model.} 
Solid arrows indicate hypothesised main effects of social, self-, and spatial presence on autonomy, competence, and relatedness. Dashed arrows capture three sets of moderating pathways: spatial presence moderating the motivational effects of social and self-presence, and gender and age moderating all presence–need associations.}}
\label{hypotheses}
\end{figure}

\qj{Building on the background, this section consolidates the theoretical grounds for our research questions and derives corresponding hypotheses. We focus first on how different dimensions of presence may relate to the satisfaction of basic psychological needs in social VR (RQ1). We then turn to boundary conditions, examining cross-dimensional effects and user-level moderators (RQ2).
\autoref{hypotheses} illustrates the overall hypothesized pathways.}

\subsection{\qj{Rationale for RQ1}}
\qj{As discussed in the previous section, a stronger presence is associated with improved task performance and motor efficiency, more fluid social coordination, and greater rapport~\cite{slater2016enhancing,skarbez2017survey}. Thus, presence, beyond the perceptual illusion of “being there,” may also function as a motivational mechanism. As SDT~\cite{ryan2017self} posits that the quality of motivation and well-being depends on satisfying three basic needs: autonomy, competence, and relatedness. If presence shapes how freely users can act, how capable they feel, and how connected they become, then different presence dimensions might map onto these needs.}

\qj{However, we know little about how multiple presence dimensions jointly shape users’ underlying psychological experiences in VR use. This omission is understandable, given that prior VR work typically examines presence in constrained laboratory settings with short-term, task-focused experimental designs with participants who knew they were in an experiment \citep{hein2018usage,Wei78}. This limits the observation of sustained, self-directed psychological satisfaction (e.g., relatedness) \citep{ryan2017self}. Understanding how autonomy, competence, and relatedness are supported in VR is crucial both for theory (clarifying the role of presence in SDT need satisfaction) and for design (informing how systems might foster healthy and intrinsically motivated participation).} 
%Existing studies in digital media and games show that presence is associated with engagement and enjoyment, but rarely clarify \emph{which} presence dimensions matter for \emph{which} psychological needs~\cite{ballou2022self,tyack2024self}.
\qj{Social VR platforms provide an opportunity to address this gap. In these environments, users’ basic psychological needs are continually negotiated through immersive, embodied social interaction. In this study, the three basic needs serve as key psychological outcomes. RQ1 asks how social, self-, and spatial presence predict these outcomes. Below we develop hypotheses for each presence dimension.}

\subsection*{\qj{Hypotheses}}

People’s perception of autonomy is influenced by feedback from their social environment. When individuals interact with their surroundings, they expect a response to their actions. 
Sufficient feedback from the environment may facilitate the perceived autonomy increases \cite{bandura1977self}. A study by \citet{jung2011understanding} in the virtual world Second Life found that social presence may positively correlate with autonomy, influencing users’ continued engagement in the virtual world. %\cite{jung2011understanding}.
Social presence may enhance a user's sense of competence by fostering performance improvements and collaborative engagement. Social Facilitation Theory \cite{zajonc1965social} suggests that the mere presence of others can enhance performance, particularly on well-learned or routine tasks. VR studies has shown that strong social presence in controlled experiments significantly affects participant performance in VR, with human social presence influencing reaction times and accuracy more than AI-driven counterparts \cite{Sutskova2023Cognitive}.  In addition, research on VR music performance found that high social presence improved coordination, synchronization, and overall musical output \cite{VanKerrebroeck2021A}. %and potentially a stronger sense of competence  

\begin{itemize}
    \item \qj{\textbf{H1a–c}: Social presence is positively associated with autonomy(a), competence(b), and relatedness(c).}
\end{itemize}

A high sense of self-presence should support all three basic needs.
%A high sense of self-presence may enhance relatedness.
%Empirical social VR studies may support this. 
\qj{Previous qualitative studies} reveals higher self-identification with avatars potentially enhances social interactions \cite{Freeman2020My, freeman2021body}.
Besides, research reveals that users often adopt self-identified avatars in social VR to express their identity, such as transgender individual. Identifying with avatars can strengthen community bonds and provide social support \cite{freeman2022re}.
%Sense of self-presence may enhance perceived competence in virtual environments.
The alignment between virtual and physical self-representation strengthens sensorimotor congruence, which has been linked to higher task efficiency and performance accuracy in VR \cite{kilteni2012sense}. %When users perceive their avatar as an extension of themselves, they are more likely to experience greater control, positively influencing perceived competence.  
Empirical research suggests that avatars closely resembling the user's physical appearance enhance self-identification, leading to greater immersion and improved spatial awareness \cite{zhao2023effects}. 
Moreover, \qj{in collaborative tasks,} users using avatars with high visual similarity to themselves demonstrated better performance and improved outcomes \cite{Land2015Does}.
Research shows that when users experience high self-presence, they feel greater agency over their movements, decisions, and interactions \cite{kilteni2012sense}. \qj{This potentially leads} to an increased sense of autonomy. This effect is particularly evident in social VR, where self-representation directly impacts how users navigate social interactions and decision-making \cite{DeVeaux2023Exploring}.
High self-presence allows users to express their identity more authentically, strengthening their sense of ownership over their virtual experience \cite{freeman2022re}.

\begin{itemize}
    \item \qj{\textbf{H2a–c:} Self-presence is positively associated with autonomy (a), competence (b), and relatedness (c).}
\end{itemize}

\qj{Prior research shows that spatial distance cues shape how people evaluate others and feel emotionally connected.}
The perceptions of physical closeness may enhance psychological and social closeness \cite{bargh2012automaticity}. Studies found that a long-distance couple using a social VR application to ``sleep next to each other'' in virtual space reported feeling emotionally closer with enhanced presence. 
Spatial presence is also associated with user autonomy. 
%Spatial presence increases when users can act freely in the environment, wich may strengthen their sense of agency~\cite{triberti2025being,Havranek2012Perspective,Herrera2006Agency}.
Spatial presence implies that users temporarily “forget” the mediation of technology, perceiving the virtual environment as a real, actionable space \cite{Sun2016Presence, wirth2007process}. This perceptual illusion of non-mediation potentially enables users to act freely and intuitively, which supports autonomous functioning in the virtual environment.  
%According to Flow Theory, competence can be strengthened when users experience deep focus and immersion in an activity \cite{nakamura2009flow}. 
A strong sense of spatial presence can deepen cognitive absorption, reduce distractions, and support task focus, which in turn enhances performance and self-efficacy \cite{Sas2003Presence, Kampling2018Feeling}. A 24-year systematic review of VR learning likewise finds that feeling ``being there'' promotes deeper cognitive involvement and higher task competence \cite{Krassmann2019What}.

\begin{itemize}
    \item \qj{\textbf{H3a–c:} Spatial presence is positively associated with autonomy (a), competence (b), and relatedness (c).}
\end{itemize}

\subsection{\qj{Rationale for RQ2}}

\qj{The preceding subsection focused on the main effects of social, self-, and spatial presence on basic psychological needs. 
However, these effects are unlikely to be uniform across situations and users. From both a theoretical and design perspective, it is important to understand when and for whom presence supports the basic needs.%, rather than assuming that higher presence is always beneficial in the same way.
}

\qj{Layered models of presence also point to such boundary conditions. 
As described by \citet{slater1997framework}, place illusion captures the perceptual sense of “being there’’ in a virtual environment and provides a basis for plausibility-related experiences of events and social interaction. 
Building on this, the integrative survey of \citet{skarbez2017survey} reviews existing models of presence and proposes a unified framework. They analyzed presence and related constructs in terms of place-related and plausibility-related aspects of experience, together with a social presence illusion component. 
On this view, spatial or place-related presence can be treated as the perceptual aspect of ``being there'' that underlies other presence-related experiences in virtual environments. 
This perspective motivates RQ2a and our hypotheses about cross-dimensional effects of spatial presence (H4).}
%More recent accounts similarly conceptualise spatial aspects of presence as a relatively low-level spatiotemporal organisation of how the self is situated in, and oriented towards, surrounding space, which can in turn scaffold more complex experiential states such as flow or narrative absorption \citep{pianzola2021presence}. 
%Social and self presence can then be seen as higher-level, content- and self-related aspects of presence that build on this basic sense of being located in the virtual environment. From this layered perspective, we treat spatial presence as a perceptual foundation that may amplify or attenuate the motivational impact of social and self presence on basic psychological needs.

\qj{At the same time, prior work shows that user characteristics such as gender and age shape both presence experiences and psychological needs 
\citep{felnhofer2012virtual,shafer2017modern,kober2014effects}. 
These characteristics can be conceptualised as between-person moderators that determine for whom particular presence dimensions support autonomy, competence, and relatedness. 
This motivates RQ2b and our hypotheses about moderation by gender and age (H5--H6).}

\subsection*{\qj{Hypotheses}}

%Building on prior theorizing, we also speculated about cross-dimensional effects. Presence models suggested that spatial presence (i.e., “being there” in a mediated environment) provides a foundation for higher-order forms such as social and self-presence \citep{slater1997framework, lee2004presence}. If spatial presence enhances the realism of co-location and embodiment, it may strengthen the motivational impact of social presence on relatedness and of self-presence on autonomy and competence.

\qj{Building on this layered view of presence outlined above, we expect spatial presence to intensify the motivational impact of other presence dimensions. When users feel more firmly ``there'' in the virtual environment, co-located others and one's avatar are likely to be experienced as more real, embodied, and actionable. Accordingly, we hypothesize that spatial presence moderate the associations of social and self-presence with autonomy, competence, and relatedness.}

\begin{itemize}
    \item \qj{\textbf{H4a--c:} Spatial presence moderates the associations between social presence and perceived autonomy (a), competence (b), and relatedness (c).}
    \item \qj{\textbf{H4d--f:} Spatial presence moderates the associations between self-presence and perceived autonomy (d), competence (e), and relatedness (f).}
\end{itemize}

Literature suggests that male and female users may differ in how they experience presence in virtual environments. For instance, \citet{felnhofer2012virtual} found that men reported significantly higher levels of spatial presence and realism than women during a virtual public speaking task. Likewise, \citet{shafer2017modern} showed that men experienced greater spatial and self-presence and reported lower cybersickness. 
\qj{Differences also appear in how they prioritize and experience psychological needs.} For example, research has shown that women often place greater emphasis on relatedness, whereas men more often prioritise competence-related experiences \cite{hartmann2006gender}. \qj{Together, these} findings suggest that gender differences may shape how presence in VR translates into basic needs. Accordingly, we propose that gender moderates the effects of presence on psychological need fulfillment.

\begin{itemize}
    \item \qj{\textbf{H5a--c:} Gender moderates the associations between social presence and perceived autonomy (a), competence (b), and relatedness (c).}
    \item \qj{\textbf{H5d--f:} Gender moderates the associations between spatial presence and perceived autonomy (d), competence (e), and relatedness (f).}
    \item \qj{\textbf{H5g--i:} Gender moderates the associations between self-presence and perceived autonomy (g), competence (h), and relatedness (i).}
\end{itemize}

Prior work suggests that age may shape how users experience immersive environments and derive psychological outcomes. For example, \citet{kober2014effects} found that older adults reported a weaker sense of presence compared to younger participants. \qj{The pattern is} attributed to factors such as lower immersive tendencies and age-related physiological or cognitive declines \cite{kober2014effects}. Beyond presence, psychological research indicates that the salience and expression of basic psychological needs also vary across the lifespan \cite{Jung2021Self}. A large cross-sectional study (ages 18–97) reported that autonomy satisfaction tends to increase with age, whereas relatedness follows a cubic trajectory \citep{lataster2022basic}.
Given these age-related dynamics in both presence and psychological needs, we hypothesize that age moderates the relationships between presence dimensions and need satisfaction.

\begin{itemize}
    \item \qj{\textbf{H6a--c:} Age moderates the associations between social presence and perceived autonomy (a), competence (b), and relatedness (c).} 
    \item \qj{\textbf{H6d--f:} Age moderates the associations between spatial presence and perceived autonomy (d), competence (e), and relatedness (f).}
    \item \qj{\textbf{H6g--i:} Age moderates the associations between self-presence and perceived autonomy (g), competence (h), and relatedness (i).}
\end{itemize}

\section{Method}
%This section describes the methodological procedures. We describe participant recruitment and sample characteristics, the measures used to assess psychological needs and presence, and the analytical procedures, including model validation and correlation analysis.
%This study combined survey-based data collection with rigorous measurement validation and structural modeling. 
We recruited participants with prior experience in social VR and measured psychological need satisfaction and sense of presence using validated scales. After validating the measurement model through \qj{Confirmatory Factor Analysis (CFA) and Exploratory Factor Analysis (EFA)}, we estimated an SEM to test hypothesized relationships, including moderation effects. The method referred to previous studies in HCI \citep{Renkai22, cai2023listen, ghaiumy2021difficulties} and aligns with recent studies on social VR that examine presence and user experience \citep{van2023feelings, barreda2022hooked}.
SEM was well suited to our aims as it simultaneously models latent constructs, accounts for measurement error, and enables the testing of multiple direct and moderating effects within a single framework \citep{kline2023principles,hoyle1995structural}.
\qj{Within this framework, our goal was not to evaluate the absolute value of each construct on any single social VR platform, but to understand structural %and cross-platform 
relationships between presence and psychological need satisfaction at the aggregated level. These relationships are modeled at the person level rather than the platform level. This approach is similar to prior studies on social media \citep{van2025online} and gaming \citep{LIAO2026108819, bian2025spectatorship}, which models psychological relationships without targeting a specific platform. And it also aligns with previous cross-platform studies of social VR \citep{barreda2022hooked,van2023feelings}.}

%at an high level rather than focusing on a specific platform
\subsection{Participants}
Data for this study were collected through Prolific, a survey panel platform widely used in academic research. Eligibility required respondents to be at least 18 years old, proficient in English, and have experience with social VR \qj{using a VR headset}. To confirm the latter, participants were provided with a short description of social VR \qj{(adapted from \citep{schulenberg2024does})}. Only those reporting prior experience were allowed to continue to the main survey. 
A total of 359 participants were recruited through Prolific. However, 58 participants were excluded from the data analysis due to such as incomplete responses or failing to meet the eligibility criteria.
\autoref{tab:1} in the \autoref{app:demo} summarizes the demographic information of the 301 valid responses for this study.

\subsection{Measurement}

\subsubsection{Self-Determination Theory}
Participants’ psychological need satisfaction in social VR was measured using adapted items from \citet{sheldon2001satisfying} and \citet{partala2011psychological}. Each basic need was assessed with three items tailored to capture experiences specific to social VR.

For example, autonomy was measured by items such as “In social VR, I feel free to do things my own way” and “I feel that my choices are based on my true interests and values” (Cronbach’s $\alpha = .825$, AVE = .615).
Competence was assessed with items such as “I feel that I am successfully completing difficult tasks and projects” and “I feel very capable in what I do” (Cronbach’s $\alpha = .878$, AVE = .707).
Relatedness was measured with items such as “I feel close and connected with other people who are important to me” and “I feel a strong sense of intimacy with the people I spend time with” (Cronbach’s $\alpha = .904$, AVE = .759).

All items were rated on a 5-point Likert scale ranging from 1 (not true at all) to 5 (completely true). The adapted scale demonstrated strong internal consistency and acceptable convergent validity across all three dimensions.

\subsubsection{Sense of Presence}

 An adapted version of the five-item social presence subscale from the Multimodal Presence Scale for virtual reality environments \citep{makransky2017development} was used to measure social presence. The scale included items such as “When I use social VR, I feel like I am in the presence of another person in the virtual environment.” The scale demonstrated excellent internal consistency (Cronbach’s $\alpha = .932$, AVE = .706).

Spatial presence was assessed using an adapted version of the four-item self-location subscale from the Spatial Presence Experience Scale \citep{hartmann2015spatial}. Items were adjusted to reflect participants’ experiences in social VR (e.g., “When I use social VR, it feels as though I am physically present in the social VR environment”). The scale showed good internal consistency (Cronbach’s $\alpha = .851$, AVE = .683).

Self-presence was measured using an adapted version of the self-presence subscale from the Multimodal Presence Scale \citep{makransky2017development}. An example item is “When I use social VR, I feel like my avatar is an extension of my real body within the virtual environment.” The scale demonstrated acceptable reliability (Cronbach’s $\alpha = .891$, AVE = .734).

All presence scales were rated on a 5-point Likert scale ranging from 1 (strongly disagree) to 5 (strongly agree). The results indicated strong internal consistency and adequate convergent validity for all presence-related constructs.

\subsection{Data Analysis}

We conducted a CFA to validate the measurement model, following \cite{brown2012confirmatory}. The results were as follows: $\chi^2 = 575.736$, $df = 215$, $\chi^2/df = 2.68$, CFI = 0.958, TLI = 0.950, and RMSEA = 0.075.
Then, we conducted an EFA. Given that our data consisted of Likert-scale responses measuring psychological perceptions, we applied the Oblimin rotation method. The EFA was performed to verify whether the expected factor structure aligned with the survey data. Item Q5-2 was originally expected to load onto the Spatial Presence factor but instead showed a stronger loading on Social Presence (>0.50), which did not align with our theoretical framework. To ensure a clearer factor structure and improve construct validity, Q5-2 was removed from further analysis.

\begin{table}[htbp]
\centering
\caption{\textbf{HTMT Ratios Among Latent Constructs.}}
\label{tab:HTMT}
\begin{tabular}{lcccccc}
\toprule
 & 1 & 2 & 3 & 4 & 5 & 6 \\
\midrule
1. Social presence     & 1.000 &       &       &       &       &       \\
2. Spatial presence    & 0.778 & 1.000 &       &       &       &       \\
3. Self presence       & 0.682 & 0.790 & 1.000 &       &       &       \\
4. Autonomy            & 0.656 & 0.591 & 0.508 & 1.000 &       &       \\
5. Competence          & 0.558 & 0.542 & 0.561 & 0.618 & 1.000 &       \\
6. Relatedness         & 0.659 & 0.610 & 0.664 & 0.592 & 0.620 & 1.000 \\
\bottomrule
\end{tabular}
\end{table}

After removing the item, the model fit improved: $\chi^2 = 468.274$, $df = 194$, $\chi^2/df = 2.41$, CFI = 0.967, TLI = 0.961, and RMSEA = 0.069, indicating a better overall fit compared to the initial model.
With a validated measurement model, we proceeded with SEM to investigate the relationships between factors. The SEM was estimated using the WLSMV estimator, considering the ordinal nature of our data. Our analysis ensured that the finalized model maintained theoretical coherence and strong psychometric properties while accurately reflecting the structure identified through EFA.
To evaluate discriminant validity, We computed the heterotrait-monotrait (HTMT) ratio of correlations, using the htmt function from the \texttt{semTools} package in R. As presented in \autoref{tab:HTMT}, all HTMT ratios are below the recommended threshold of 0.85, indicating that the variables demonstrate adequate discriminant validity.

We further explored the relationships between variables using Pearson correlation analysis, the results of which are presented in \autoref{tab:Pe}. All variables exhibited statistically significant correlations with one another at the 99\% confidence level.

\begin{table}[htbp]

\caption{\textbf{Pearson Correlation Analysis Among Factors. Note: ** $p < .01$. All correlations are significant at the 1\% level.}}
\begin{tabular}{@{}lllllll@{}}
\toprule
                      & 1     & 2     & 3     & 4     & 5     & 6     \\ \midrule
1. Social presence    & 1.000 &  &  &  &  &  \\
2. Spatial   presence & 0.860** & 1.000 & &  &  & \\
3. Self   presence    & 0.765** & 0.860** & 1.000 & & &  \\
4. Autonomy           & 0.750** & 0.671** & 0.583** & 1.000 & &  \\
5. Competence         & 0.652** & 0.625** & 0.635** & 0.701** & 1.000 & \\
6. Relatedness        & 0.744** & 0.685** & 0.722** & 0.681** & 0.702** & 1.000 \\ \bottomrule
\end{tabular}
\label{tab:Pe}
\end{table}

To test moderation effects, we adopted different strategies based on the nature of the moderator variables. Spatial presence, treated as a continuous latent construct, was tested using latent interaction modeling within SEM (H4). This approach allowed us to examine how spatial presence intensifies or attenuates the effects of social and self presence on psychological needs. \qj{In line with standard practice for latent variable interactions \citep{maslowsky2015estimating,klein2000maximum}, social, self-, and spatial presence were retained as main-effect predictors while also entering the model via latent interaction terms.}

\begin{table*}[htbp]
\caption{\qj{\textbf{SEM results for direct effects.}
Note: Estimates are fully standardized coefficients ($\beta$, Std.all).
$^*$ $p<.05$, $^{**}$ $p<.01$, $^{***}$ $p<.001$; n.s. = non-significant.}}
\centering
\begin{tabular}{p{0.50\textwidth} c p{0.18\textwidth}}
\hline
\qj{\textbf{Hypothesis (paths)}} & \qj{\textbf{Estimate ($\beta$)}} & \qj{\textbf{Support}} \\
\hline
\qj{\textbf{H1(a–c): Social presence $\rightarrow$ need satisfaction}} & & \qj{\textbf{Full support}} \\
\hspace{1em}\qj{(a) Social presence $\rightarrow$ Relatedness}  & \qj{0.40$^{***}$}  & \qj{\ding{51}~(H1a)} \\
\hspace{1em}\qj{(b) Social presence $\rightarrow$ Competence}   & \qj{0.30$^{**}$}   & \qj{\ding{51}~(H1b)} \\
\hspace{1em}\qj{(c) Social presence $\rightarrow$ Autonomy}     & \qj{0.49$^{***}$}  & \qj{\ding{51}~(H1c)} \\
\hline
\qj{\textbf{H2(a–c): Self-presence $\rightarrow$ need satisfaction}} & & \qj{\textbf{Partial support}} \\
\hspace{1em}\qj{(a) Self-presence $\rightarrow$ Autonomy}       & \qj{0.02 (n.s.)}   & \\
\hspace{1em}\qj{(b) Self-presence $\rightarrow$ Competence}     & \qj{0.30$^{**}$}   & \qj{\ding{51}~(H2b)} \\
\hspace{1em}\qj{(c) Self-presence $\rightarrow$ Relatedness}    & \qj{0.41$^{***}$}  & \qj{\ding{51}~(H2c)} \\
\hline
\qj{\textbf{H3(a–c): Spatial presence $\rightarrow$ need satisfaction}} & & \qj{\textbf{No support}} \\
\hspace{1em}\qj{(a) Spatial presence $\rightarrow$ Competence}  & \qj{0.07 (n.s.)}   & \\
\hspace{1em}\qj{(b) Spatial presence $\rightarrow$ Autonomy}    & \qj{0.19 (n.s.)}   & \\
\hspace{1em}\qj{(c) Spatial presence $\rightarrow$ Relatedness} & \qj{-0.03 (n.s.)}  & \\
\hline
\end{tabular}
\label{tab:SEMD_summary}
\end{table*}

In contrast, gender and age were treated as categorical grouping variables, and moderation was tested using multi-group SEM (H5–H6). Gender was coded based on self-identification (male vs. female), excluding participants with low-frequency gender identities due to statistical power concerns\footnote{Participants who identified as non-binary or third gender (n = 6) were not included in the gender-based multi-group SEM analysis due to insufficient sample size. While this exclusion was necessary for methodological reasons, we acknowledge the importance of representing gender diversity in virtual environments and encourage future work to oversample or specifically focus on underrepresented identities to enable more inclusive analyses.}. Age was split at 35 years: a commonly used developmental threshold and consistent with prior research on using age as the moderating role \citep[e.g.,][]{liebana2014antecedents,liebana2017factors}. This threshold provided sufficient sample sizes in both groups while capturing meaningful psychological distinctions across age cohorts. 

\begin{table*}[htbp]
\caption{\qj{\textbf{SEM results for the moderating role of spatial presence.}
Note: Estimates are standardized coefficients ($\beta$). 
$^*$ $p<.05$, $^{**}$ $p<.01$, $^{***}$ $p<.001$; n.s. = non-significant.}}
\centering
\begin{tabular}{p{0.60\textwidth} c p{0.18\textwidth}}
\hline
\qj{\textbf{Hypothesis (paths)}} & \qj{\textbf{Estimate ($\beta$)}} & \qj{\textbf{Support}} \\
\hline
\qj{\textbf{H4a--c: Spatial presence $\times$ social presence $\rightarrow$ need satisfaction}} & & \qj{\textbf{No support}} \\
\hspace{1em}\qj{(a) Social presence $\times$ spatial presence $\rightarrow$ Competence}   & \qj{$-0.039$ (n.s.)} & \\
\hspace{1em}\qj{(b) Social presence $\times$ spatial presence $\rightarrow$ Autonomy}     & \qj{$0.002$ (n.s.)}  & \\
\hspace{1em}\qj{(c) Social presence $\times$ spatial presence $\rightarrow$ Relatedness}  & \qj{$-0.057$ (n.s.)} & \\
\hline
\qj{\textbf{H4d--f: Spatial presence $\times$ self-presence $\rightarrow$ need satisfaction}} & & \qj{\textbf{No support}} \\
\hspace{1em}\qj{(d) Self-presence $\times$ spatial presence $\rightarrow$ Competence}     & \qj{$0.082$ (n.s.)}  & \\
\hspace{1em}\qj{(e) Self-presence $\times$ spatial presence $\rightarrow$ Autonomy}       & \qj{$0.007$ (n.s.)}  & \\
\hspace{1em}\qj{(f) Self-presence $\times$ spatial presence $\rightarrow$ Relatedness}    & \qj{$0.036$ (n.s.)}  & \\
\hline
\end{tabular}
\label{tab:SEMM_summary}
\end{table*}

\section{Results}

This section presents the \qj{analysis results}. We first report the main effects of the three presence dimensions on psychological needs (H1–H3). Next, we demonstrate the moderating role of spatial presence (H4), followed by multi-group SEM analyses examining moderation by gender (H5) and age (H6).

\subsection{Main Effects of Presence Dimensions}

Once the measurement model was validated, we used SEM to test the hypotheses H1 to H3. 
The structural equation model showed excellent fit to the data (\(\chi^2(194) = 287.40\), CFI = 0.997, TLI = 0.996, RMSEA = 0.040, SRMR = 0.050).

\qj{The results indicate differences in how the three presence dimensions relate to psychological need satisfaction (see Table~\ref{tab:SEMD_summary}).
Social presence emerged as the consistent predictor across all three needs.}
%Regarding the effect of Social Presence, our SEM results fully support H1(a), H1(b), and H1(c). %Specifically, higher social presence significantly predicted greater perceived autonomy ($\beta = 0.494, p < .001$), competence ($\beta = 0.295, p = .006$), and relatedness ($\beta = 0.398, p < .001$), supporting \textbf{H1(a–c)}.
\qj{Self-presence positively predicted competence and relatedness but showed no significant association with autonomy.}
%Self Presence, our results partially support H2. While self-presence significantly predicted competence ($\beta = 0.301, p = .003$) and relatedness ($\beta = 0.413, p < .001$), it did not significantly predict autonomy ($p = .835$), hence \textbf{H2(b) and H2(c)} are \textbf{supported}, but \textbf{H2(a) is not supported}.
\qj{Finally, spatial presence did not show any significant predictive effects on the three needs.} %Thus, \textbf{H3(a), H3(b), and H3(c) are not supported}.
%These findings indicate that Social Presence is the strongest and most consistent predictor of psychological needs in social VR, with Self Presence also contributing to Competence and Relatedness, but not Autonomy. 
%Table \ref{tab:SEMD} summarizes all SEM results, including the level of support for each hypothesis.

\subsection{Moderating Effect of Spatial Presence}

We further conducted SEM to test whether spatial presence moderates the relationships between social presence, self presence, and psychological needs. The fit indices for the moderation model indicate an excellent fit: $\chi^2(197) = 484.619$, CFI = 0.924, TLI = 0.910, RMSEA = 0.070, SRMR = 0.064.

\qj{\autoref{tab:SEMM_summary} presents the results.} As shown, none of the interaction effects are statistically significant. This suggests that spatial presence does not moderate the relationships between social presence, self presence, and the psychological needs of autonomy, competence, and relatedness  (see  \autoref{fig:moderation_social} and \autoref{fig:moderation_self} in \autoref{app:plots} \qj{for the visualization of the interaction effect)}.

\begin{table*}[htbp]
\centering
\caption{\qj{\textbf{Gender moderation: multi-group SEM results.} 
\textit{Note.} $\Delta\chi^2$ values represent chi-square difference tests comparing 
a model constraining the focal path to equality across gender groups with a model 
freely estimating the path (df = 1). Significant $\Delta\chi^2$ values indicate a 
moderating effect by gender. Coefficients are standardized ($\beta$). 
$^* p<.05$, $^{**} p<.01$, $^{***} p<.001$.}}
\label{tab:gender_moderation_summary}
\resizebox{\textwidth}{!}{%
\begin{tabular}{p{0.54\textwidth} c c c c p{0.16\textwidth}}
\hline
\qj{\textbf{Hypothesis (paths)}} & \qj{$\boldsymbol{\Delta\chi^2}$} & \qj{\textbf{Sig.}} & 
\qj{\textbf{Male ($\beta$)}} & \qj{\textbf{Female ($\beta$)}} & \qj{\textbf{Results}} \\
\hline
\qj{\textbf{H5a--c: Gender $\times$ social presence $\rightarrow$ need satisfaction}} & & & & & 
\qj{\textbf{Partial support}} \\
\hspace{1em}\qj{(a) Social presence $\rightarrow$ Competence}   & \qj{1.97}  & \qj{n.s.}      & \qj{0.074}    & \qj{0.446}    & \\
\hspace{1em}\qj{(b) Social presence $\rightarrow$ Autonomy}     & \qj{10.86} & \qj{$^{***}$}  & \qj{0.462}    & \qj{1.185}    & \qj{\ding{51}~(H5b)} \\
\hspace{1em}\qj{(c) Social presence $\rightarrow$ Relatedness}  & \qj{4.79}  & \qj{$^*$}      & \qj{0.215}    & \qj{0.686}    & \qj{\ding{51}~(H5c)} \\
\hline
\qj{\textbf{H5d--f: Gender $\times$ spatial presence $\rightarrow$ need satisfaction}} & & & & &
\qj{\textbf{Partial support}} \\
\hspace{1em}\qj{(d) Spatial presence $\rightarrow$ Competence}  & \qj{1.39}  & \qj{n.s.}      & \qj{$-0.102$} & \qj{0.207}    & \\
\hspace{1em}\qj{(e) Spatial presence $\rightarrow$ Autonomy}    & \qj{39.60} & \qj{$^{***}$}  & \qj{0.413}    & \qj{$-0.575$} & \qj{\ding{51}~(H5e)} \\
\hspace{1em}\qj{(f) Spatial presence $\rightarrow$ Relatedness} & \qj{0.86}  & \qj{n.s.}      & \qj{0.099}    & \qj{$-0.175$} & \\
\hline
\qj{\textbf{H5g--i: Gender $\times$ self-presence $\rightarrow$ need satisfaction}} & & & & &
\qj{\textbf{Partial support}} \\
\hspace{1em}\qj{(g) Self-presence $\rightarrow$ Competence}     & \qj{5.63}  & \qj{$^*$}      & \qj{0.569}    & \qj{0.133}    & \qj{\ding{51}~(H5g)} \\
\hspace{1em}\qj{(h) Self-presence $\rightarrow$ Autonomy}       & \qj{0.23}  & \qj{n.s.}      & \qj{$-0.152$} & \qj{$-0.028$} & \\
\hspace{1em}\qj{(i) Self-presence $\rightarrow$ Relatedness}    & \qj{0.13}  & \qj{n.s.}      & \qj{0.433}    & \qj{0.329}    & \\
\hline
\end{tabular}%
}
\label{tab:gender_moderation}
\end{table*}

\begin{table*}[htbp]
\centering
\caption{\qj{\textbf{Age moderation: multi-group SEM results.}
\textit{Note.} $\Delta\chi^2$ values represent chi-square difference tests comparing
a model constraining the focal path to equality across age groups with a model
freely estimating the path (df = 1). Significant $\Delta\chi^2$ values indicate a
moderating effect by age. Coefficients are standardized ($\beta$).
$^* p<.05$, $^{**} p<.01$, $^{***} p<.001$.}}
\label{tab:age_moderation}

\renewcommand{\arraystretch}{1.05}

\resizebox{\textwidth}{!}{%
\begin{tabular}{p{0.52\textwidth} c c c c p{0.16\textwidth}}
\hline
\qj{\textbf{Hypothesis (paths)}} & \qj{$\boldsymbol{\Delta\chi^2}$} & \qj{\textbf{Sig.}} &
\qj{\textbf{Under 35 ($\beta$)}} & \qj{\textbf{Over 35 ($\beta$)}} & \qj{\textbf{Support}} \\
\hline
\qj{\textbf{H6a--c: Age $\times$ social presence $\rightarrow$ need satisfaction}}
& & & & & \qj{\textbf{Partial support}} \\[-2pt]
\hspace{1em}\qj{(a) Social presence $\rightarrow$ Competence}   & \qj{0.30} & \qj{n.s.} & \qj{0.212} & \qj{0.518} & \\
\hspace{1em}\qj{(b) Social presence $\rightarrow$ Autonomy}     & \qj{0.39} & \qj{n.s.} & \qj{0.305} & \qj{0.773} & \\
\hspace{1em}\qj{(c) Social presence $\rightarrow$ Relatedness}  & \qj{4.24} & \qj{$^*$} & \qj{0.493} & \qj{0.186} & \qj{\ding{51} (H6c)} \\
\hline
\qj{\textbf{H6d--f: Age $\times$ spatial presence $\rightarrow$ need satisfaction}}
& & & & & \qj{\textbf{No support}} \\[-2pt]
\hspace{1em}\qj{(d) Spatial presence $\rightarrow$ Competence}  & \qj{2.28} & \qj{n.s.} & \qj{0.206} & \qj{$-0.255$} & \\
\hspace{1em}\qj{(e) Spatial presence $\rightarrow$ Autonomy}    & \qj{0.88} & \qj{n.s.} & \qj{0.218} & \qj{$-0.023$} & \\
\hspace{1em}\qj{(f) Spatial presence $\rightarrow$ Relatedness} & \qj{0.00} & \qj{n.s.} & \qj{0.142} & \qj{0.119}    & \\
\hline
\qj{\textbf{H6g--i: Age $\times$ self-presence $\rightarrow$ need satisfaction}}
& & & & & \qj{\textbf{Partial support}} \\[-2pt]
\hspace{1em}\qj{(g) Self-presence $\rightarrow$ Competence}     & \qj{0.26} & \qj{n.s.} & \qj{0.229} & \qj{0.402}   & \\
\hspace{1em}\qj{(h) Self-presence $\rightarrow$ Autonomy}       & \qj{6.01} & \qj{$^*$} & \qj{0.166} & \qj{$-0.147$} & \qj{\ding{51} (H6h)} \\
\hspace{1em}\qj{(i) Self-presence $\rightarrow$ Relatedness}    & \qj{2.26} & \qj{n.s.} & \qj{0.158} & \qj{0.512}   & \\
\hline
\end{tabular}%
}
\end{table*}

\subsection{Moderating Effect of Gender}

To examine whether gender moderates the relationships \qj{between the three presence dimensions on psychological outcomes}, we conducted a multi-group SEM analysis using the \texttt{lavaan} package in R. Participants were grouped by gender (Male: $n = 189$, Female: $n = 106$), and a free multi-group model was estimated to allow all parameters to vary across groups.
The unconstrained multi-group model showed good fit to the data: $\chi^2(388) = 618.28$, $p < .001$; CFI = 0.93; TLI = 0.92; RMSEA = 0.07 [90\% CI: 0.055, 0.072]; SRMR = 0.063.

To test moderation, each structural path was constrained to be equal across groups and compared against the unconstrained model using chi-square difference tests.

\qj{As shown in Table~\ref{tab:gender_moderation}, four paths were moderated by gender out of nine. For females, social presence more strongly predicted autonomy and relatedness; for males, spatial presence more strongly supported autonomy, and self-presence more strongly supported competence. These findings suggest gendered differences in how presence predicts psychological needs.}

\subsection{Moderating Effect of Age}

To investigate whether age moderates the relationships between \qj{presence dimensions and psychological outcomes}, we conducted a multi-group SEM analysis based on age groups (Under 35: $n = 202$, Over 35: $n = 99$).

The unconstrained multi-group SEM model yielded good fit statistics: $\chi^2(388) = 652.89$, $p < .001$; CFI = 0.92; TLI = 0.90; RMSEA = 0.067 [90\% CI: 0.059, 0.076]; SRMR = 0.062.

To test age as a moderator, we imposed equality constraints on each structural path and conducted chi-square difference tests against the freely estimated model. A significant chi-square difference suggests that the effect is moderated by age.

\qj{As shown in Table~\ref{tab:age_moderation}, two paths were significantly moderated by age. Self-presence more strongly predicted autonomy among younger participants. In addition, social presence was more strongly associated with relatedness in the under-35 group. This indicates age-dependent variation in how social and self-related presence cues support basic psychological needs.}

\section{Discussion}
\qj{This study examined how social, self-, and spatial presence predict the satisfaction of autonomy, competence, and relatedness in social VR. Two key insights emerge. First, the three presence dimensions map onto these psychological needs in distinct ways. It reveals that presence in social VR also functions as a motivational mechanism more than a perceptual state. Second, these presence–need links are not universal but moderated by user characteristics, including gender and age. It indicates that the motivational impact of presence depends on who the users are. 
Together, these findings refine and extend prior work on presence and motivation in VR environments.
The remainder of this section discusses these results within the literature and their implications in three steps. First, we unpack the differential motivational roles of presence (Section~\ref{sec:presence_needs}). We then examine how these pathways vary by gender and by age (Section~\ref{sec:individual_differences}). Finally, we discuss the implications of these patterns (Section~\ref{sec:implica}).}
\final{Importantly, the discussion offers plausible explanations that are consistent with our findings and prior literature. Given the correlational and the possibility of unmeasured confounds, they should not be read as definitive accounts.}

%rather than certainties about underlying mechanisms.

\subsection{\qj{The Differential Motivational Roles of Presence in Need Satisfaction}}
\label{sec:presence_needs}

Our results provide strong support for the idea that different dimensions of presence play distinct roles in satisfying users’ basic psychological needs in social VR. This extends prior literature on presence and motivation.

\subsubsection{Social Presence as a Core Driver}

\qj{Our results show that social presence is a strong and consistent predictor of autonomy, competence, and relatedness}. 

\qj{Social presence is especially important for relatedness. It fits prior qualitative work in social VR showing that feeling genuinely “with others” supports emotional expression~\citep{chen2025understanding}, intimacy~\citep{Hugging5}, and social bonding~\citep{piitulainen2022vibing}.
Social presence also contributes meaningfully to autonomy and competence.} This suggests that social interaction itself can function as an empowering context, where feelings of freedom and mastery may emerge through meaningful engagement with others. Whereas traditional SDT perspectives tend to emphasize personal agency as the root of autonomy, our findings enhance the view that autonomy can be socially constructed \cite{deci2000and}. \qj{In this sense, social presence functions as a psychological enabler that channels interpersonal experiences into feelings of choice and efficacy. }

\qj{Compared to spatial and self presence, which have been traditionally highlighted} for their roles in immersion and identity enactment, our findings underscore a distinct contribution: social presence is the most functionally integrative dimension of presence in motivating psychological need satisfaction. \qj{This positions social presence as a central motivational pathway in social VR, rather than a perceptual sense of co-location.} This advances theoretical understanding of the differentiated roles presence dimensions play in immersive media. 

\subsubsection{Self-Presence as a Selective Contributor}
Compared to social presence, self-presence appears to play a more selective motivational role in shaping psychological outcomes. Our results show that self-presence significantly predicts perceived competence and relatedness, but not autonomy. 

For competence, self-presence may help users feel more in control of their actions in a virtual space. When users feel that their avatar moves in ways that match their own intentions and body movements, they are more likely to feel confident and skilled. %This effect has been found in online gaming and task-focused virtual embodiment \citep{kilteni2012sense}. % Our findings suggest that self-presence fosters competence, possibly by reinforcing the feeling of embodied control (i.e., users feel “in command” of their actions, which supports their perceived efficacy).
In terms of relatedness, self-presence allows the avatar to function as an extension of the social self. \qj{Previous qualitative studies in social VR suggest that users' emotional expression, identity communication, and social bonding become more natural and authentic when users identify with their avatar \citep{freeman2022re, Freeman2020My}.} %, particularly in communities such as LGBTQ+ or furry fandoms. 
It offers users a medium for self-affirmation and group belonging. Our quantitative results further support this perspective: self-presence significantly predicts relatedness satisfaction. \qj{It shows self-presence's role in facilitating the cognitive-emotional transition from simply being in the environment to feeling genuinely connected to others.}

However, \qj{the absence of a significant link to autonomy is noteworthy.} Even when users feel embodied and capable, they may not feel truly self-directed. One reason could be the limited behavioral flexibility current avatar systems provide: while basic movements are tracked in real time, many expressive actions (e.g., facial expressions) still rely on preset animations \cite{kolesnichenko2019understanding}. \qj{Users may therefore experience control over execution, but not over the authorship of their actions.} %In other words, they experience control without agency, a condition that could be described as \textit{functional control} rather than  \textit{intentional authorship}.
This fits with self-efficacy theory \citep{bandura1977self}, which separates the feeling of being capable from the feeling of being in control of specific goals.

\subsubsection{Spatial Presence as a Perceptual Backdrop}

\final{In social VR,} spatial presence did not significantly predict any of the three basic psychological needs and showed no significant moderating effects in combination with social or self presence (i.e., H3 and H4 were not supported). 
This lack of direct or interactive influence suggests that spatial presence may play a more supportive role in motivational processes rather than serving as an active contributor.

\qj{At first glance, this may seem counterintuitive, given that spatial immersion is often treated as a core feature of VR.} However, it may indicate that mere immersion or “being there” is not sufficient to satisfy deeper psychological needs in the absence of engaging social or self-relevant content. Spatial presence likely functions as an environmental anchor \qj{that connects “me” to the virtual space and provides a realistic context for action. But by itself, the anchor does not generate feelings of autonomy, competence, or social bonding.} In other words, a user can feel present in a visually impressive virtual room, but unless they have meaningful control or social interaction in that room, their basic needs may remain unfulfilled. 
\qj{A recent workplace VR study for knowledge workers points in the same direction \citep{ppali2025vr}. Short, self-selected VR breaks that combine different well-being activities within stylized environments improved mindfulness and reduced anxiety. Participants value appropriate activities and atmosphere more than visual realism.}
This perspective aligns with arguments that spatial presence may be a necessary but not sufficient condition. Users likely \qj{require a basic} feeling of “being there” before higher-order social presence can emerge \citep{oh2018systematic}, \qj{but once that baseline is reached,} social and self-related factors dominate the impact on user experience.   
\qj{In this sense, spatial presence is better conceptualized as a perceptual backdrop that enables, but does not itself activate, motivational outcomes.}

Our null findings for moderation effects further support this interpretation. Spatial presence did not significantly moderate the effects of social or self presence on basic psychological need satisfaction. This may be because spatial presence acts less as a conventional moderator and more as a perceptual threshold, a gatekeeper that enables but does not amplify other effects.
Once users cross a minimal spatial presence threshold, the psychological impact of social and self-presence can unfold. Similar “switch-type” variables have been discussed in psychological conditions (e.g., baseline trust, perceived control) as prerequisites rather than moderators \cite{
wang2016diffusion,papadopoulou2001trust, otten2023paving}. Thus, spatial presence may best be understood as a perceptual backdrop that supports motivational outcomes but does not interact directly with them in a statistical sense.
\qj{\final{Taken together, our findings do not support a direct-effect account of spatial presence in predicting need satisfaction.} Instead, spatial presence provides the backdrop on top of which social and self presence can operate as primary motivational pathways.}

\subsection{\qj{Individual Differences in Presence–Need Pathways}}
\label{sec:individual_differences}

\qj{Our results show that presence–need pathways vary by gender and age. 
\final{These patterns point to differences in the strength of presence--need relationships across groups, highlighting which presence dimensions are most predictive of need satisfaction for which users in our sample.}}

\subsubsection{Gender Differences in Presence Effects on Psychological Need Satisfaction}
\label{sec:gender_moderation}

%\qj{Aligning with the second key insight, 
\qj{The results show} that four presence--need pathways were moderated by gender: (1) social presence → autonomy, (2) social presence → relatedness, (3) spatial presence → autonomy, and (4) self-presence → competence.

Social presence supported autonomy and relatedness more strongly for women than for men. \qj{This pattern suggests that the presence of others in VR plays a particularly important role in women’s need fulfilment}. It may relate to gender differences in social sensitivity and relational motivation. Research has shown that women are more likely to perceive social feedback and internalize it as validation of self-worth and motivational support \cite{xia2023moderating}.
\qj{It is noteworthy that these tendencies may themselves also be shaped by sociocultural context. As women’s social roles in many contexts place a stronger emphasis on relational care, emotional attunement, and maintaining harmony \citep{verma2025women,brody2000gender}.}
Besides, self-presence predicted competence more strongly for men than for women. \qj{Feeling identified with and embodied in the avatar appears to more directly enhance men’s sense of capability and control.} \qj{Prior research in VR and gaming contexts has found that men, especially when using self-resembling avatars, tend to show} higher performance and self-efficacy \cite{kleinlogel2024effect}. In contrast, female users may feel more self-conscious or constrained by avatar representations \cite{kleinlogel2024effect}, particularly when facing stereotype threat or lacking personalization. %They have been shown to reduce perceived competence.  

Spatial presence was also more strongly related to autonomy for men. This may reflect gender-based differences in how users interpret and navigate spatial environments. Men may perceive spatial control and exploration as reinforcing their sense of autonomy, while women may rely more heavily on social and contextual cues to construct agency. 
\qj{Again, cultural factors likely play a role. In many gaming and VR cultures that have historically been male-dominated, navigating and “conquering” virtual environments is framed as a valued form of skill and independence \citep{condis2018gaming}. This framing may make spatial immersion particularly autonomy-relevant for men.}
\qj{Prior work suggests that men tend to report higher spatial presence \citep{yoon2015user}. \final{Our findings go further by suggesting that, in our sample, spatial presence is more strongly associated with autonomy-related need satisfaction for men.}}

\subsubsection{Age Differences in Presence–Need Satisfaction Pathways}
\label{sec:age_moderation}

\qj{Beyond gender, this study found that the links between social presence and relatedness, and between self-presence and autonomy,} were significant only among younger participants. These results indicate that age shapes how presence is translated into basic psychological need satisfaction.

Younger users were more responsive to social presence in fulfilling their relatedness needs. 
\qj{Younger individuals are generally more accustomed to digital social interaction} with messaging apps, online games, or social media integrated into daily life \citep{pew2008games,kontos2010communication}. Prior research suggests that younger adults tend to maintain larger and more dynamic social networks \citep{Lansford1998Satisfaction, bobzien2025visualizing}, often supported through digital platforms. \qj{This familiarity may make it easier for them to treat virtual social cues as authentic and emotionally meaningful. As a result, feeling that others are “there” in VR more readily translates into a sense of connection for younger users.} 
In contrast, older adults tend to emphasize emotional depth and familiarity in their relationships. They often prioritize a smaller circle of close ties over broader online networks \citep{hulur2020rethinking, Kim2020Connecting,yu2018facebook}. \qj{For older participants, we did not observe the same strength of association between social presence and relatedness; relatedness may depend more on whether interactions are meaningful, sustained, or linked to pre-existing relationships.} 
Self-presence was positively associated with autonomy among younger users, but not among older users. One explanation is that \qj{younger individuals are more used to treating avatars as tools for self-expression and agency.} In digital games and virtual environments, they frequently explore alternative identities and roles through avatars \citep{kafai2007your,fox2013avatars}. Such practices highlight their capacity to choose and enact one’s preferred self, which may strengthen the perception of autonomy. 
Older users are less likely to perceive avatars as direct extensions. They often face difficulties in identifying with humanlike avatars, partly due to lower familiarity with avatar-based interaction and a stronger reliance on real-world identities \citep{cheong2011avatar}. Consequently, \final{the link between avatar embodiment and self-concept is weaker,} and avatars contribute less to autonomy for older users.
\final{However, because relevant experience such as gaming was not measured, we cannot exclude the possibility that the observed differences partly reflect differences in prior exposure rather than the demographic features per se.}
%\final{As noted above, these age differences may also be confounded with unmeasured gaming/VR experience, which future work should assess directly.}

\subsection{\qj{Implications for the Design of Immersive Environments}}
\label{sec:implica}

%\subsection{Conceptual and Practical Implications for Immersive Design}

\qj{In this section, we organise the practical implications for immersive environments around two key takeaways we previously outlined. First, we highlight the implications of understanding presence dimensions from a motivational perspective. Then we discuss what it means for design that these pathways vary across user groups. }

\subsubsection{\qj{Linking Presence Dimensions to Motivational Pathways}}

\qj{Our findings suggest that presence can be understood in motivational terms \final{in the social VR context.} In this view, social presence operates as a core motivational driver, self-presence makes a more selective contribution, and spatial presence functions largely as a perceptual backdrop that scaffolds other processes.} %Rather than treating presence as a uniform “more is better” experience of being there, our results point to differentiated roles for each presence dimension in supporting users’ motivation.

\qj{This perspective helps extend functional-chain accounts that underlie much prior work on immersive design. For example, many studies have identified design features that enhance presence \citep{Cummings2016ImmersionPresence} (e.g., more expressive avatars, higher graphical fidelity). And other work has examined downstream consequences of heightened presence, typically in terms of performance outcomes \citep{palombi2023role,wei2025towards} (e.g., learning, training, attention, and cognition or task in general). Taken together, a substantial portion of the research can be broadly characterized by a functional logic of the form: features → presence → performance. 
In contrast, %building on prior evidence for the features → presence link, 
our results provide a complementary motivational chain: features → presence → needs → motivation. By showing that different presence dimensions are differentially associated with need satisfaction, our study enriches the explanatory scope of presence research beyond perceptual or performance metrics.}

\qj{This complementary chain has several practical implications. Extensive VR research and design work has focused on increasing specific forms of presence \citep{Cummings2016ImmersionPresence,wei2025towards}, often under the (sometimes implicit) assumption that stronger presence yields higher engagement. 
 However, these efforts have rarely examined whether such enhancements actually strengthen motivated use. By explicitly linking presence dimensions to distinct need-satisfaction pathways, our study addresses this gap and clarifies which forms of presence are most relevant for sustained motivation.
This perspective makes earlier feature-focused findings more actionable for motivational design. Prior work has extensively shown that certain features will reliably enhance particular forms of presence.
When designers know both (a) which features reliably enhance a given type of presence and (b) which presence dimensions, in turn, support basic psychological needs, they can use these features more deliberately to build VR environments that foster engagement and sustained use.}

%Moreover, this pathway perspective makes earlier feature-focused findings more actionable for motivational design. Prior work has shown that certain features reliably boost social or spatial presence, but often stopped at presence as the final outcome. Our results show that in our sample, increases in social presence are associated with greater need satisfaction and intrinsic motivation. When designers know both (a) which features reliably enhance a given type of presence and (b) which presence dimensions, in turn, support basic psychological needs, they can use these features more deliberately to build social VR environments that foster engagement and sustained use.

\qj{Beyond these general implications, this perspective also has direct relevance for domain-specific immersive systems. Rather than attempting to maximize an undifferentiated sense of presence, designers should pay particular attention to the specific presence dimensions most relevant to their intended outcomes.
For example, for applications seeking to enhance training, where a sense of competence is important \citep{lofgren2017effects}. Our results indicate that investing in features that promote rich social presence and self-presence will be effective.
%Supporting self-presence is particularly valuable when autonomy or competence are central.
By contrast, our findings caution against relying on spatial fidelity alone as an enhancement. Enhanced graphical realism or immersion should be viewed as an enabling condition rather than a direct lever for need satisfaction. In practice, this means that once a reasonable threshold of “being there” is achieved (e.g. comfortable field of view, good fidelity), additional effort is better spent on content and features that make the environment meaningful rather than merely more realistic. }

\qj{These implications also extend to evaluation practices. Assessing immersive systems should move beyond global presence scores. They should also assess how different presence dimensions, and their supporting design features, contribute to distinct aspects of need satisfaction. When adding a new world, test whether users feel recognized, effective, and connected, not only on whether the environments are more realistic. }

%Taken together, this integrative approach can inform the design of social VR experiences by directing attention to conditions under which users feel present with others and as themselves, rather than merely in a place. Prioritising social and self-presence while treating spatial presence as an enabling backdrop makes it more likely that technological immersion will translate into psychologically need-satisfying and motivating experiences.

\subsubsection{\qj{Tailoring Presence Pathways to Diverse Users}}
\qj{A second key theme is around presence--motivation pathways are not universal. 
Our moderation results show that user characteristics shape how different forms of presence translate into basic psychological needs. In our sample, women appeared to derive more need satisfaction from social presence, whereas men benefited more from self-presence and spatial presence. Younger users more readily converted social presence into relatedness and self-presence into autonomy, while older users did not experience these gains automatically.}

\qj{These patterns argue against a one-size-fits-all approach to presence design. For users whose need satisfaction is more contingent on social presence (which, in our sample, more often included women), features that support emotionally safe, responsive, and meaningful interaction are likely to be particularly important. For users whose need satisfaction is more closely tied to self- and spatial presence (more common among men in our sample), providing precise avatar control, responsive embodiment, and goal-relevant navigation may better leverage these pathways.
For younger users, maximising social presence (e.g., larger lobbies, rich communication tools) and avenues for avatar personalisation may be key to fulfilling relatedness and autonomy needs.
For older users, however, simply intensifying immersion or social density is unlikely to be sufficient. Developers might instead focus on facilitating meaningful encounters through, for example, smaller gatherings, structured activities, or features that support continuity with existing offline relationships.}

\qj{Rather than prescribing separate designs for different demographic groups, we see these patterns as evidence that platforms should offer configurable combinations of social, self-, and spatial features. This allows users with different gendered and age-related experience profiles to translate presence into autonomy, competence, and relatedness in ways that fit them. A well-designed platform might, for instance, offer both collaborative social environments (for those who crave interaction and relatedness) and sandbox modes or skill-based mini-games (for those who enjoy independent mastery and competence).  
In such situations, additional design layers, such as matchmaking and recommendation, can further support the fit. For example, if a segment of users tends to benefit more from social presence, recommendation algorithms could preferentially route them to rooms or content, where responsiveness is easier to achieve. }%If men tend to benefit more from activity-oriented, agency-rich contexts, recommendations could emphasise skill-based or action-oriented spaces in which embodied control and exploration are central.

\qj{The emergence of VR applications for specific user groups, such as platforms tailored to older adults \citep{baragash2022virtual}, will likely continue. Our findings can inform the design and evaluation for sustaining intrinsic motivation in such contexts. Rather than only asking whether users “like” a feature, designers and evaluators can ask whether it enhances the specific forms of presence that matter for their users’ need satisfaction. 
In targeted contexts, for instance, this includes questions such as: Does this feature make interpersonal cues easier to notice and interpret? More generally, claims that an environment is “for a particular user group” can be examined empirically by assessing whether its concrete features strengthen the most important presence for them and, in turn, support autonomy, competence, and relatedness.}

\subsection{Limitations and Future Works}
\qj{While this study offers novel insights into the motivational effects of presence, several limitations should be noted.}
First, although we examined moderation by factors such as gender and age, other moderators, such as personality traits, technological familiarity, \final{or relevant prior experience (e.g., gaming)} were not included. \qj{We also did not directly measure cultural background; region of residence was collected only as a demographic indicator and was skewed toward Europe and North America.} 
These factors may also shape how presence dimensions relate to motivational outcomes, and their \qj{omission means that some variance likely remains unexplained.} 
In addition, while the sample size was sufficient for the analyses, it was somewhat skewed toward younger participants and showed a gender imbalance, with more male than female respondents. 
\qj{In addition, all constructs were assessed using retrospective self-report questionnaires that asked respondents to recall their general experiences in social VR (“when I use social VR …”). This approach is common in large-scale survey research and has been used to measure presence in both social VR \citep{barreda2022hooked, van2023feelings} and other online mediated environments such as virtual worlds and online gaming \citep{BEHMMORAWITZ2013119, de2007digital}. These studies similarly relied on participants’ habitual or typical experiences rather than momentary responses following a specific exposure. %It differs from laboratory studies of presence that administer instruments immediately after a specific VR exposure. 
However, in contrast to the typical session-based presence assessment paradigm, the approach may be more prone to being influenced by memory and context effects. 
%However, retrospective assessments may be influenced by memory and context effects \citep{schwarz2001asking}. 
These issues are especially relevant for our age-moderation analyses, as younger and older adults may differ in how they be able to recall their prior VR experiences. Our findings should therefore be interpreted as associations between typical (rather than momentary) levels of presence and psychological need satisfaction in social VR, and age-related findings should be interpreted with caution. } 
%Future research could employ diary or experience-sampling methods that collect presence ratings directly after individual VR sessions, and explicitly compare such session-level measures with retrospective summary ratings across different age groups

Building on the current study’s findings and limitations, several directions emerge for future research to advance both the theory and practical application of presence and motivation in social VR.
1) This study used SEM to test hypothesized relationships. Future research should adopt longitudinal designs to assess causality and dynamic change. For example, longitudinal tracking could reveal whether the effects of presence on psychological need satisfaction strengthen or fade over time as users become more familiar with the virtual environment. 
2) While this study focused on basic needs from SDT, future research could examine additional outcomes that are highly relevant in immersive environments. These include social identity and belongingness, particularly in communities that rely on avatar representation (e.g., LGBTQ+ groups). They also include affective outcomes such as emotion regulation, enjoyment, and empathy in response to presence-enhancing features.
\qj{3) Our models are specified at the person level across social VR platforms. A natural next step is to build on these relationships with studies that either focus on a single well-characterised platform or systematically compare different design variants (e.g., avatar systems, interaction tools) to derive more feature-specific implications.}

\section{Conclusion}

\qj{This paper examined how three types of presence in social VR affect the satisfaction of people’s basic psychological needs, using Self-Determination Theory as a guiding framework. The results suggest that presence is a psychologically multifaceted construct with distinct functional roles. Social presence was the strongest and most consistent predictor of all three needs. Self-presence contributed more selectively, supporting competence and relatedness but not autonomy. Spatial presence showed no direct or moderating effects, suggesting its role as a perceptual backdrop rather than directly driving motivation.} Beyond those, the study uncovered moderating influences of gender and age. Women benefited more from social presence for autonomy and relatedness, whereas men gained more from self- and spatial presence, especially for competence and autonomy. \qj{Younger users showed stronger links between social presence and relatedness, and between self-presence and autonomy, than older users.} These results underscore that presence effects are shaped by user characteristics, challenging assumptions of uniform psychological benefits from immersive design features. 
Theoretically, \qj{this research extends presence theory by framing presence not only as perceptual realism but as a motivational mechanism that supports psychological needs in different ways.} Practically, the findings offer insights for designing more inclusive and psychologically supportive social VR systems. %Features that enhance real-time social engagement, identity expression, and embodied control can be strategically implemented to meet diverse user needs.

%Taken together, this study provides a multidimensional, demographically sensitive account of how presence functions in immersive environments, giving insights for future work that explores the interplay between presence, motivation, and human experience in the evolving landscape of virtual interaction.  

\section*{\qj{Use of Generative AI Tools}}
\qj{The authors occasionally used a generative AI tool to improve the grammar and language clarity. The tool was not used to generate ideas, conduct data analysis, or create scientific content. All scientific contributions, including the study design, analyses, and interpretation of the results, were developed and verified by the authors.}

\begin{acks} 
\final{We thank the anonymous reviewers for their constructive and thoughtful feedback, which helped us significantly improve the paper. This work was supported by the Research Council of Finland (Grant \#357270).}
\end{acks}

\bibliographystyle{ACM-Reference-Format}
\bibliography{sample-base}

\appendix

\section{Demographics}
\label{app:demo}

\autoref{tab:1} presents the demographic information of the 301 valid survey respondents, including age, gender, frequency of usage, and region of residence.

\begin{table}[H]
\begin{tabular}{@{}lp{2.5cm}cc@{}}
\toprule
\qj{Variable}            & \qj{Response Category}                     & Frequency & Percent \\ \midrule
Age                 & 18-24 years old           & 58        & 19.2    \\
                    & 25-34 years old           & 145       & 48.2    \\
                    & 35-44 years old           & 53        & 17.6    \\
                    & 45-54 years old           & 32        & 10.6    \\
                    & 55-64 years old           & 12        & 4       \\
                    & 65+ years old             & 1         & 0.3     \\
Gender              & Male                      & 189       & 62.8    \\
                    & Female                    & 106       & 35.2    \\
                    & Non-binary/ third gender & 6         & 2       \\
Frequency of Usage  & Once a month or less      & 107       & 35.5    \\
                    & A few times a month       & 109       & 36.2    \\
                    & A few times a week        & 66        & 21.9    \\
                    & Daily                     & 19        & 6.3     \\
Residence by Region & Europe                    & 179       & 59.5    \\
                    & North America             & 81        & 26.9    \\
                    & South America             & 5         & 1.7     \\
                    & Africa                    & 24        & 8       \\
                    & Oceania                   & 1         & 0.3     \\
                    & Asia                      & 11        & 3.7     \\ \bottomrule
\end{tabular}
\caption{\textbf{Demographic information of the 301 valid responses for this study}}
\label{tab:1}
\end{table}

\section{Interaction Plots}
\label{app:plots}

\begin{figure}[H] 
    \centering
    \includegraphics[width=1\linewidth]{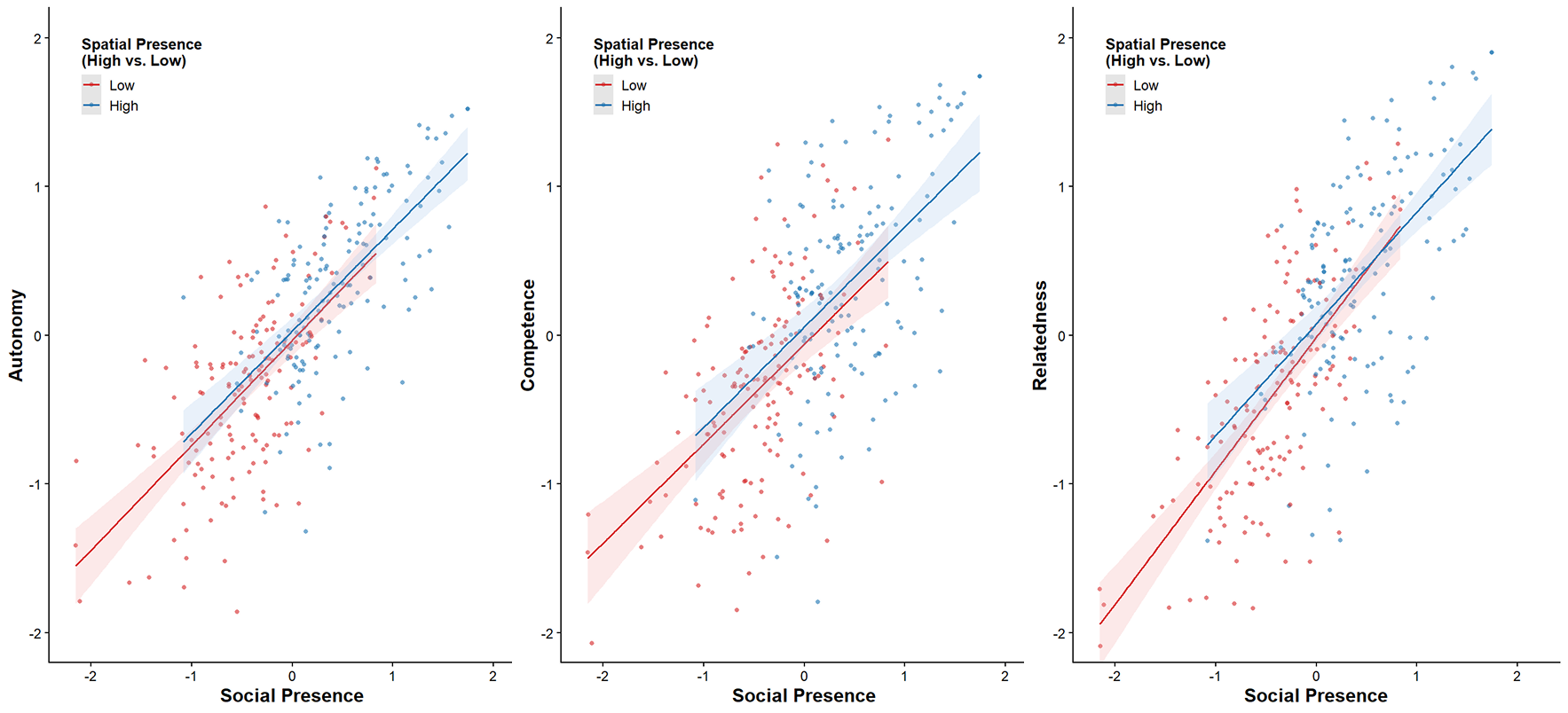}
    \caption{\qj{\textbf{Interaction effects between social presence and spatial presence on basic psychological needs.} (Left) Autonomy, (Middle) Competence, and (Right) Relatedness are plotted as a function of social presence, separately for low (red) and high (blue) spatial presence. Shaded bands represent 95\% confidence intervals.}}  
    \label{fig:moderation_social}
\end{figure}

\begin{figure}[H]
    \centering
    \includegraphics[width=1\linewidth]{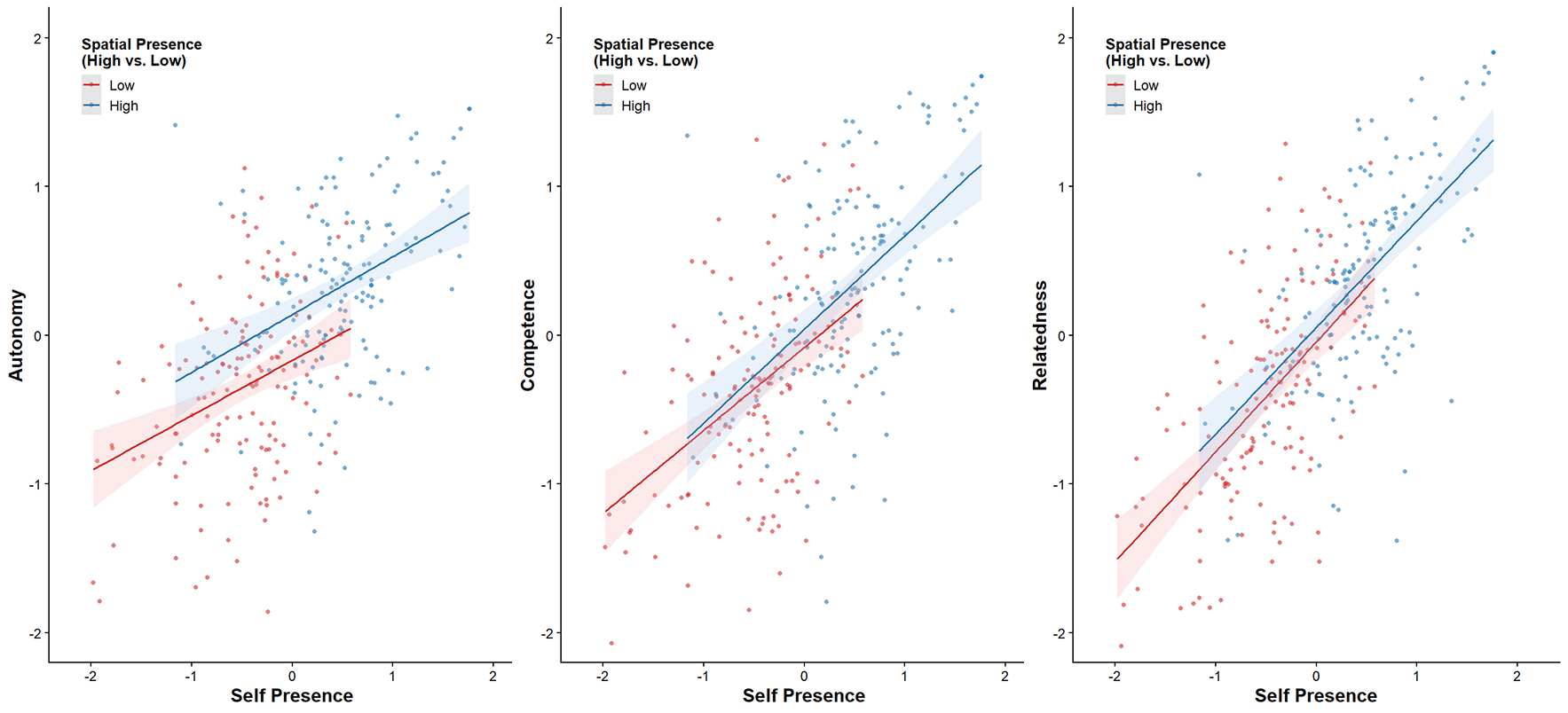}
    \caption{\qj{\textbf{Interaction effects of Self Presence and Spatial Presence on the three basic psychological needs.} (Left) Autonomy, (Middle) Competence, and (Right) Relatedness are plotted as a function of Self Presence, separately for low (red) and high (blue) Spatial Presence. Shaded bands represent 95\% confidence intervals.}}
    \label{fig:moderation_self}
\end{figure}

Figures~\ref{fig:moderation_social} and~\ref{fig:moderation_self} display the tested interactions between spatial presence and social/self-presence on autonomy, competence, and relatedness.

\end{document}